\documentclass[aps,prb,reprint,superscriptaddress,longbibliography]{revtex4-2}

\usepackage[T1]{fontenc}
\usepackage[utf8]{inputenc}
\usepackage{times,amsmath,amssymb,graphicx,enumitem}
\usepackage[dvipsnames]{xcolor}
\usepackage[colorlinks=true,allcolors=Blue]{hyperref}

\usepackage{xspace}
\usepackage{multirow}
\usepackage{float}

\usepackage{fancyhdr}
\setcitestyle{numbers, square}

\begin{document}

 \newcommand{\ie}{i.e.,\xspace}
 \newcommand{\eg}{e.g.,\xspace}
 \newcommand{\etal}{{\it et al}.\xspace}

 \newcommand{\RuCl}{$\alpha$-RuCl$_3$}

 \newcommand{\Bc}{$B_{c}$}
 \newcommand{\Bstar}{$B^{\star}$}

 \newcommand{\TN}{$T_{N}$}

 \newcommand{\Kxx}{$\kappa_{xx}$}
 \newcommand{\Kxy}{$\kappa_{xy}$}







\title{Oscillations in the magnetothermal conductivity of $\boldsymbol{\alpha}$-RuCl$_3$:
Evidence of transition anomalies
}
\author{\'Etienne~Lefran\c{c}ois}
\email{etienne.lefrancois@usherbrooke.ca}
%
\affiliation{Institut Quantique, D\'epartement de physique \& RQMP, Universit\'e de Sherbrooke, Sherbrooke, Qu\'ebec, Canada}
\author{Jordan~Baglo}
\email{jordan.baglo@usherbrooke.ca}
\thanks{\'E.L. and J.B. contributed equally.}
\affiliation{Institut Quantique, D\'epartement de physique \& RQMP, Universit\'e de Sherbrooke, Sherbrooke, Qu\'ebec, Canada}
\author{Q.~Barth\'elemy}
\affiliation{Institut Quantique, D\'epartement de physique \& RQMP, Universit\'e de Sherbrooke, Sherbrooke, Qu\'ebec, Canada}
\author{S.~Kim}
\affiliation{Department of Physics, University of Toronto, Toronto, Ontario, Canada}
\author{Young-June~Kim}
\affiliation{Department of Physics, University of Toronto, Toronto, Ontario, Canada}

\author{Louis~Taillefer}
\email{louis.taillefer@usherbrooke.ca}
\affiliation{Institut Quantique, D\'epartement de physique \& RQMP, Universit\'e de Sherbrooke, Sherbrooke, Qu\'ebec, Canada}
\affiliation{Canadian Institute for Advanced Research, Toronto, Ontario, Canada}


\date{\today}


\begin{abstract}

The 2D layered insulator~\RuCl~is a candidate material for a quantum spin-liquid state, 
which may be realized when a magnetic field 
suppresses the antiferromagnetic order present at low temperature.
%
%
Oscillations in the field dependence of the thermal conductivity, observed for an in-plane magnetic field $B$ up to a critical field \Bstar, 
have been attributed to exotic charge-neutral fermions, viewed as evidence of a quantum spin-liquid state between
the critical field~\Bc~$\simeq~7$~T~at which the antiferromagnetic phase ends and~\Bstar{}.
%
Here we report measurements of the thermal conductivity of~\RuCl~as a function of magnetic field up to 15 T applied in two distinct in-plane directions: parallel and perpendicular to the Ru-Ru bond. 
We find that the number of oscillations between~\Bc~and~\Bstar~is the same for the two field directions even though the field interval between~\Bc~and~\Bstar~is different. 
In other words, the period of the oscillations is controlled by the transition fields~\Bc~and~\Bstar. 
We conclude that these are not true oscillations---coming from putative fermions in a spin-liquid state---but anomalies associated with a sequence of magnetic transitions.

\end{abstract}

\maketitle


\section{INTRODUCTION}

\begin{figure}[H]\centering
    \includegraphics[width = 0.39\textwidth]{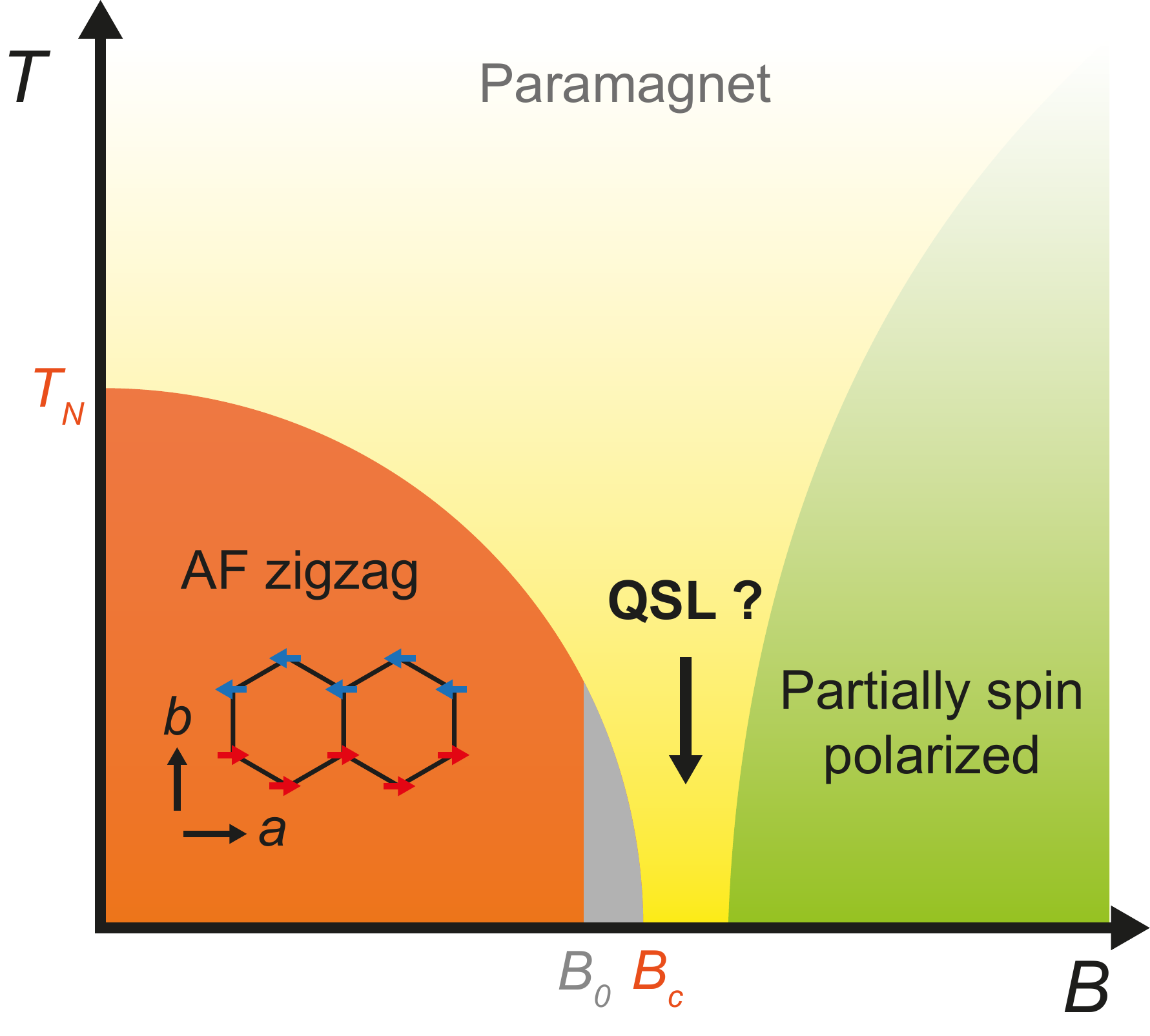}
    \caption{
    Phase diagram of \RuCl~as a function of temperature $T$ and in-plane magnetic field $B$.
    %
    The phase of long-range antiferromagnetic (AF) order is shown in orange,
    with a 
    transition temperature \TN~$\simeq{}7$~K and
    a 
    critical field \Bc~$\simeq{}7$~T.
    Spins on the Ru sites are arranged in a zigzag pattern of ferromagnetic chains (see sketch),
    but that pattern changes at $B_0{}\simeq{}6$~T, just before reaching \Bc.\
    %
    The question is whether the state just above the AF phase is a quantum spin liquid (QSL).
    }
    \label{Fig1}
\end{figure}

A major objective in the field of quantum materials is to confirm experimentally, in a real material, the existence of a quantum spin-liquid (QSL) state predicted theoretically.
In this respect, the magnetic insulator~\RuCl~is a promising material~\cite{jackeli_mott_2009,plumb_rucl3_2014,johnson_monoclinic_2015}, with its Ru atoms lying on weakly coupled 2D honeycomb layers whose interactions are such that they nearly satisfy the Kitaev model~\cite{kitaev_anyons_2006}, a model whose exact solution is a QSL with Majorana fermions as emergent excitations.
In reality, the ground state of~\RuCl~is not a QSL, but a state with long-range antiferromagnetic order,
setting in below a critical temperature~\TN~$\simeq~7$~K (Fig.~\ref{Fig1}).
However, this ordered state can be suppressed by applying a magnetic field parallel to the honeycomb layers,
until it ends at a critical field~\Bc~$\simeq~7$~T (Fig.~\ref{Fig1}).
The question then is this: What is the nature of the state just above~\Bc, at low temperature?

To address that question experimentally, thermal transport has emerged as a fruitful probe.
Early measurements of the thermal Hall conductivity, \Kxy, revealed the existence of a non-zero \Kxy, which was attributed to the Majorana fermions expected from the Kitaev model, given indications of a plateau in \Kxy~vs field at a half-quantized value~\cite{kasahara_majorana_2018}.
Although some later studies again find indications of half-quantization~\cite{yamashita_sample_2020,yokoi_half-integer_2021,bruin_robustness_2022},
others do not, and instead attribute the measured \Kxy~to chiral magnons~\cite{czajka_planar_2022} or phonons~\cite{lefrancois_evidence_2022,hentrich_large_2019}.
It is fair to say that based on \Kxy~data the case for a QSL in \RuCl~above \Bc~is currently not compelling.

\begin{figure}[H]\centering
    \includegraphics[width = 0.45\textwidth]{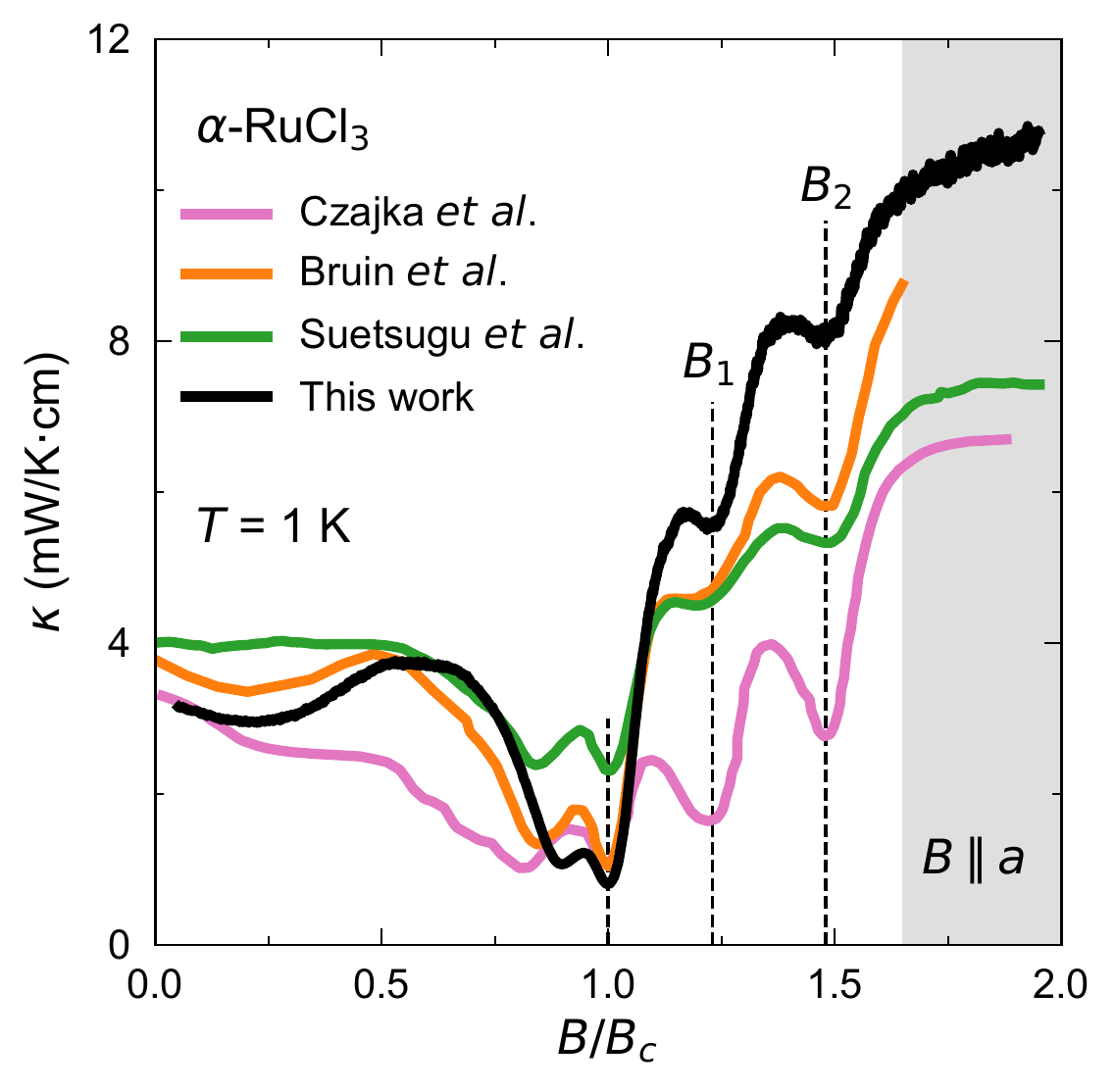}
    \caption{
    Thermal conductivity of \RuCl~as a function of in-plane magnetic field $B$, at $T~\simeq~1$~K,
    plotted as $\kappa$~vs $B/$\Bc, where \Bc~is the critical field where the antiferromagnetic phase ends (Fig.~\ref{Fig1}).
    Data from four different studies are compared:
    Czajka \etal (pink; $T = 0.96$~K and \Bc~$= 7.21$~T)~\cite{czajka_oscillations_2021};
    Bruin \etal (orange; $T = 1.0$~K and \Bc~$= 7.20$~T)~\cite{bruin_origin_2022};
    Suetsugu \etal (green; $T = 1$~K and \Bc~$= 7.15$~T)~\cite{suetsugu_evidence_2022};
    this work (black, sample S1; $T = 1.16$~K and \Bc~$= 7.69$~T).
    In all cases, the field is applied along the $a$ axis
    and the current is parallel to the field ($J \parallel B)$.
    The vertical dashed lines mark the location of three minima, at \Bc, $B_1$, and $B_2$.
    These minima in the oscillations of $\kappa$ vs $B$ are seen to be in the same locations for all four studies.
    The grey shaded region marks the regime at high field where oscillations are no longer observed.}
\label{Fig2}
\end{figure}

In parallel with thermal Hall studies, measurements of the longitudinal thermal conductivity, \Kxx~(or $\kappa$),
have revealed the existence of oscillations as a function of in-plane magnetic field $B$~\cite{czajka_oscillations_2021,bruin_origin_2022,suetsugu_evidence_2022} (see Fig.~\ref{Fig2}).
These oscillations have been interpreted as quantum oscillations akin to those produced by Landau quantization of electron states 
in a metal when a magnetic field is applied, but this time coming from putative charge-neutral fermions, {\it e.g.} gapless spinons with a Fermi surface.
Emergent neutral fermions would be a clear signature of a QSL state, albeit a different one from that expected from the Kitaev model~\cite{kitaev_anyons_2006,villadiego_pseudoscalar_2021}.
While other groups confirm the existence of these oscillations in $\kappa$~vs $B$ (Fig.~\ref{Fig2}), they attribute them to a sequence of magnetic transitions~\cite{bruin_origin_2022,suetsugu_evidence_2022}.
In this paper, we revisit these oscillations, with a focus on their anisotropy as the field direction within the honeycomb plane
is changed from being perpendicular to the Ru-Ru bond ($B \parallel a$) to being parallel ($B  \parallel b$) (Fig.~\ref{Fig1}).


\section{METHODS}

Single crystals were grown via the chemical vapor transport (CVT) method, using RuCl$_3$ powder from Sigma-Aldrich. 
The powder, composed of 45--55\% ruthenium, was sealed in a quartz ampoule under vacuum and the ampoule was then placed inside a two-zone tube furnace.
The powder was annealed for two days at $\sim\!800~^\circ$C in a temperature gradient of 70~$^\circ$C (warmest side was 850~$^\circ$C), followed by a cooldown at 4~$^\circ$C/hour while maintaining the temperature gradient.
For more details, see Refs.~\onlinecite{sears_phase_2017,kim_rucl3_2022}.
Here we report data on two as-grown (uncut) crystals, labeled S1 and S2, handled very carefully to minimize any strain induced
when installing the contacts and fixing them on the experimental mount.
Samples S1 and S2 are rectangular platelets with planar surface area 1~mm~$\times$~1~mm and thicknesses of 130 and 110~$\mu$m, respectively.
The contacts on the samples were made by attaching 25~$\mu$m-diameter silver wires with DuPont 4929N silver paste.
The heater was attached to the sample with a 100~$\mu$m-diameter silver wire with silver paste as well.

Measurements were performed by a steady-state method using a standard four-terminal technique, with the thermal current applied along the length of the sample within the honeycomb layers: perpendicular to the Ru-Ru bonds for S1 ($J \parallel a$) and parallel to the Ru-Ru bonds for S2 ($J \parallel b$).
The thermal conductivity~$\kappa$~was measured by employing a standard one-heater--two-thermometers method, using a 10~k$\rm \Omega$~resistor and two RuOx chip sensors whose magnetoresistances have been carefully taken into account:
for interpolation of temperature from measured resistance and field, detailed $R(T,B)$ calibration surfaces were collected for each individual thermometer \textit{in situ}.
All thermometers were measured using Lake Shore model 370 temperature controllers.
A main calibrated RuOx sensor placed in the field-compensated mixing chamber region of the dilution refrigerator was used for the reference temperature~$T_0$, with typical control stability within $\pm$6~$\mu$K at 100 mK and better than 0.03\% over the full measurement range.
A constant heat current~$J$~was injected at one end of the sample, while the other end of the sample is heat sunk to a copper block held at a temperature~$T_0$.
The heat current was generated by sending an electric current through the 10 k$\rm \Omega$ strain gauge,
whose resistance was measured to be essentially independent of temperature and magnetic field.
The longitudinal temperature gradient~$\Delta T = T^+ - T^-$~is measured at two points along the length of the sample, separated by a distance~$l$.
The longitudinal thermal conductivity is given by~$\kappa = J / \left( \Delta T \alpha \right)$, where~$\alpha = w t / l$~is the geometric factor of the sample (width~$w$, thickness~$t$).

The two temperatures $T^+$ and $T^-$ were measured as the magnetic field, applied parallel to the honeycomb layers, was slowly swept from 0 to 15~T,
while keeping the temperature $T_0$ constant.
For each sample, a series of field sweeps, taken at various temperatures $T_0$ from 0.1 to 5 K, were obtained
for two field directions: along the $a$~axis and along the $b$~axis of the crystal structure (Fig.~\ref{Fig1}).
The magnetic field was swept sufficiently slowly (0.04~T/min) to avoid any magnetocaloric effect within the sample during the measurement, verified to be minimal by comparison with sweeps at different rates.
The samples were firmly attached to their respective heat sinks in order to withstand the torque that could result from any slight misalignment
of the field away from the intended high-symmetry direction, and thus avoid any bending of the samples.
The error bars on the absolute values of $\kappa$ come mostly from the uncertainty in estimating the sample dimensions ($l$, $w$,~and~$t$), 
amounting approximately to $\pm$20\%.
The applied current was chosen such that~$\Delta T / T \simeq$~3--5~\%
and the resulting~$\kappa$~was verified to be independent of applied current.



\begin{figure}[t]\centering
    \includegraphics[width = 0.4\textwidth]{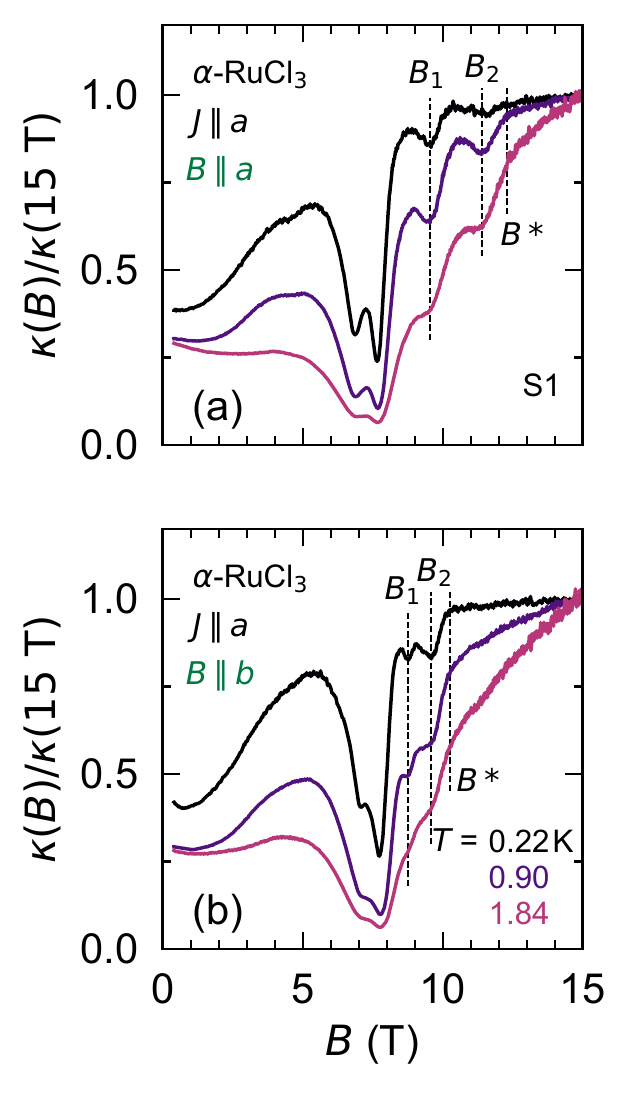}
    \caption{
    Normalized thermal conductivity 
    of sample S1 at~$T$~= 0.22, 0.88, and 1.84~K, plotted as~$\kappa$($B$)/$\kappa$(15~T) vs $B$,
    for a thermal current~$J$~applied along the $a$~axis.
    The magnetic field was applied parallel to the $a$~axis (a)  and perpendicular to~the $a$~axis (b).}
    \label{Fig3}
\end{figure}



\begin{figure}[t]\centering
    \includegraphics[width = 0.4\textwidth]{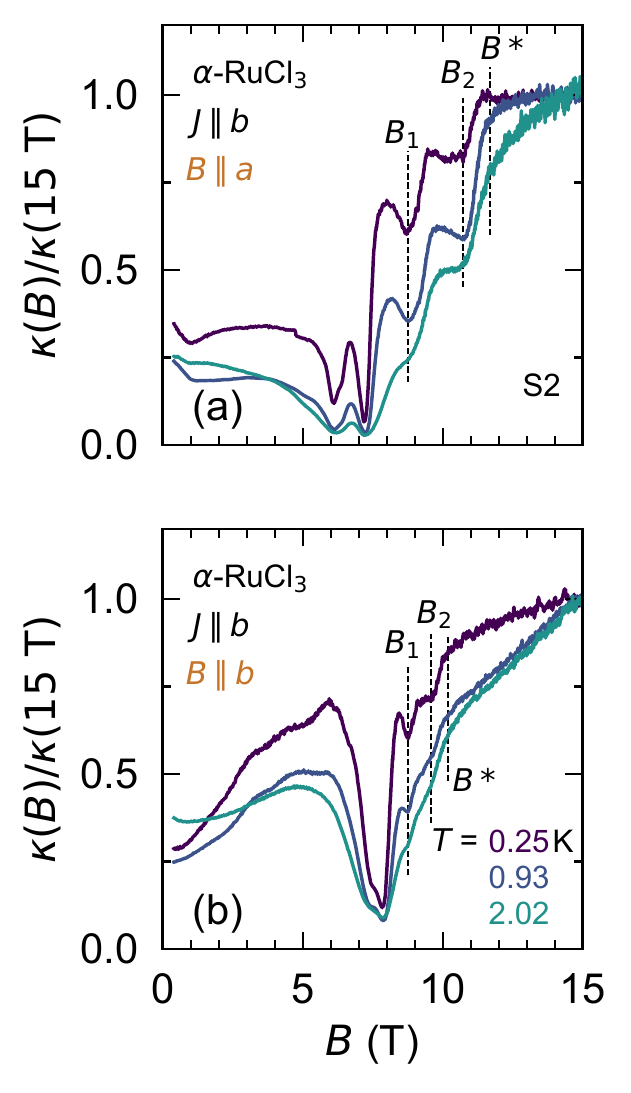}
    \caption{
    Normalized thermal conductivity 
    of sample S2 at~$T$~= 0.24, 0.90, and 2.02~K, plotted as~$\kappa$($B$)/$\kappa$(15~T) vs $B$,
    for a thermal current~$J$~applied along the $b$~axis.
   The magnetic field was applied perpendicular to the $b$~axis (a) and parallel to the $b$~axis (b).}
\label{Fig4}
\end{figure}



\section{RESULTS}

In Fig.~\ref{Fig3}a, we display three isotherms of $\kappa(B)$ taken on sample S1 ($J \parallel a$) with $B \parallel a$, plotted as $\kappa(B) / \kappa$(15~T) vs $B$. 
We observe four minima. 
The deepest minimum, located at \Bc~$\simeq 7.6$~T, corresponds to the end of the AF phase, as determined by thermodynamic measurements such as specific heat~\cite{bachus_thermodynamic_2020,suetsugu_evidence_2022}, magnetostriction~\cite{schonemann_thermal_2020,kocsis_magnetoelastic_2022} and magnetization~\cite{bruin_origin_2022}.
The minimum located at a field $B_0 \simeq 6.5$~T, just below \Bc, marks a transition internal to the AF phase (Fig.~\ref{Fig1}) where the AF ordering changes from one spin pattern to another~\cite{balz_field-induced_2021}.
The upper two minima, at $B_1$ and $B_2$, are both above \Bc, and so outside the phase of bulk AF order.
The location of $B_1$ and $B_2$ is nicely consistent with prior studies, as seen in Fig.~\ref{Fig2}.
It is possible that weak signatures of structure in $\kappa(B)$ are visible further below $B_0$ and in the AF state, in all four studies; this may be related to similar small anomalies in dielectric constant observed by Mi \textit{et al.} \cite{mi_stacking_2021}.
The main purpose of our study was to investigate the anisotropy of this oscillatory pattern.
In Fig.~\ref{Fig3}b, we display the corresponding isotherms taken on the same sample (S1) but for a field direction in the other high-symmetry in-plane direction, namely $B \parallel b$.
We again observe only two minima above \Bc, but this time they are much closer to each other (and to \Bstar).
Indeed, the separation between $B_1$ and $B_2$, which one might view as the ``period'' of the oscillations ($\Delta B \equiv B_2 - B_1$), is roughly twice as large for $B \parallel a$ compared to $B \parallel b$.

In order to confirm this anisotropy in the period of the oscillations as the in-plane field changes direction, we measured a second sample (S2), this time with the heat current flowing along the $b$~axis ($J \parallel b$).
These data are displayed in Fig.~\ref{Fig4}, where we see that two minima are present above \Bc~and their separation $\Delta B$ is again
significantly larger for $B \parallel a$ compared to $B \parallel b$.
The current direction does not seem to make a significant difference.

An important feature in our data is the existence of a threshold field, \Bstar, above which there are no oscillations (or minima) anymore.
This is especially clear in the isotherms at the lowest temperature ($T \simeq 0.2$~K), as in 
Fig.~\ref{Fig3}b for $B \parallel b$, where \Bstar~$\simeq 10$~T, and in Fig.~\ref{Fig4}a for $B \parallel a$, where \Bstar~$\simeq 12$~T.
Note the anisotropy of \Bstar.

\begin{figure}[t]\centering
    \includegraphics[width = 0.48\textwidth]{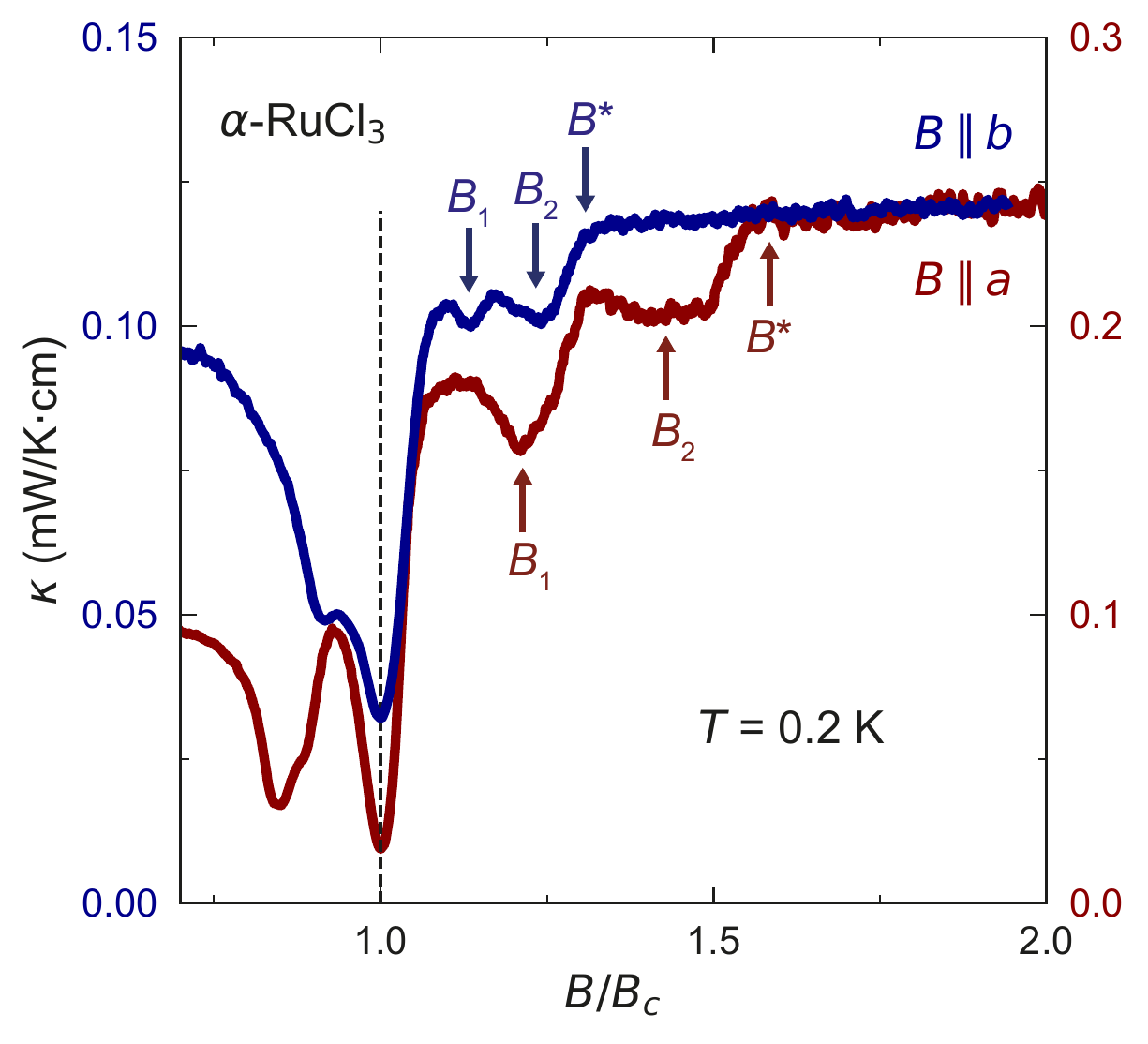}
    \caption{
    Anisotropy of characteristic fields. 
    Comparison of isotherms at $T \simeq 0.2$~K, plotted as $\kappa$ vs $B/$\Bc, for the two field directions:
    $B \parallel a$ (red; sample S2, $T = 0.20$~K, \Bc~$= 7.2$~T) 
    vs
    $B \parallel b$ (blue; sample S1, $T = 0.22$~K, \Bc~$= 7.7$~T).
    Arrows mark the location of the two minima at $B_1$ and $B_2$, and the threshold field \Bstar~above which oscillations are no longer observed.
    The period of the oscillations ($| B_1 - B_2 |$) is seen to scale with the width of the interval between the field-induced transitions ($|$\Bc~$-$~\Bstar$|$), 
    both shrinking roughly by half when the field is redirected from
    $B \parallel a$ to $B \parallel b$.
    This suggests that the oscillations are related to the transitions.}
    \label{Fig5}
\end{figure}

In Fig.~\ref{Fig5}, we directly compare isotherms for $B \parallel a$ and $B \parallel b$, both at $T \simeq 0.2$~K.
We see that the two oscillations observed for both field directions, with minima at $B_1$ and $B_2$, fit neatly in the interval between the two transitions, at \Bc~(end of AF phase) and \Bstar~(end of oscillatory pattern), even though that interval shrinks by a factor of 2 in going from $B \parallel a$ to $B \parallel b$.
This reveals a connection between oscillations and transitions, the main finding of our anisotropy study.

\begin{figure}[t]\centering
    \includegraphics[width = 0.48\textwidth]{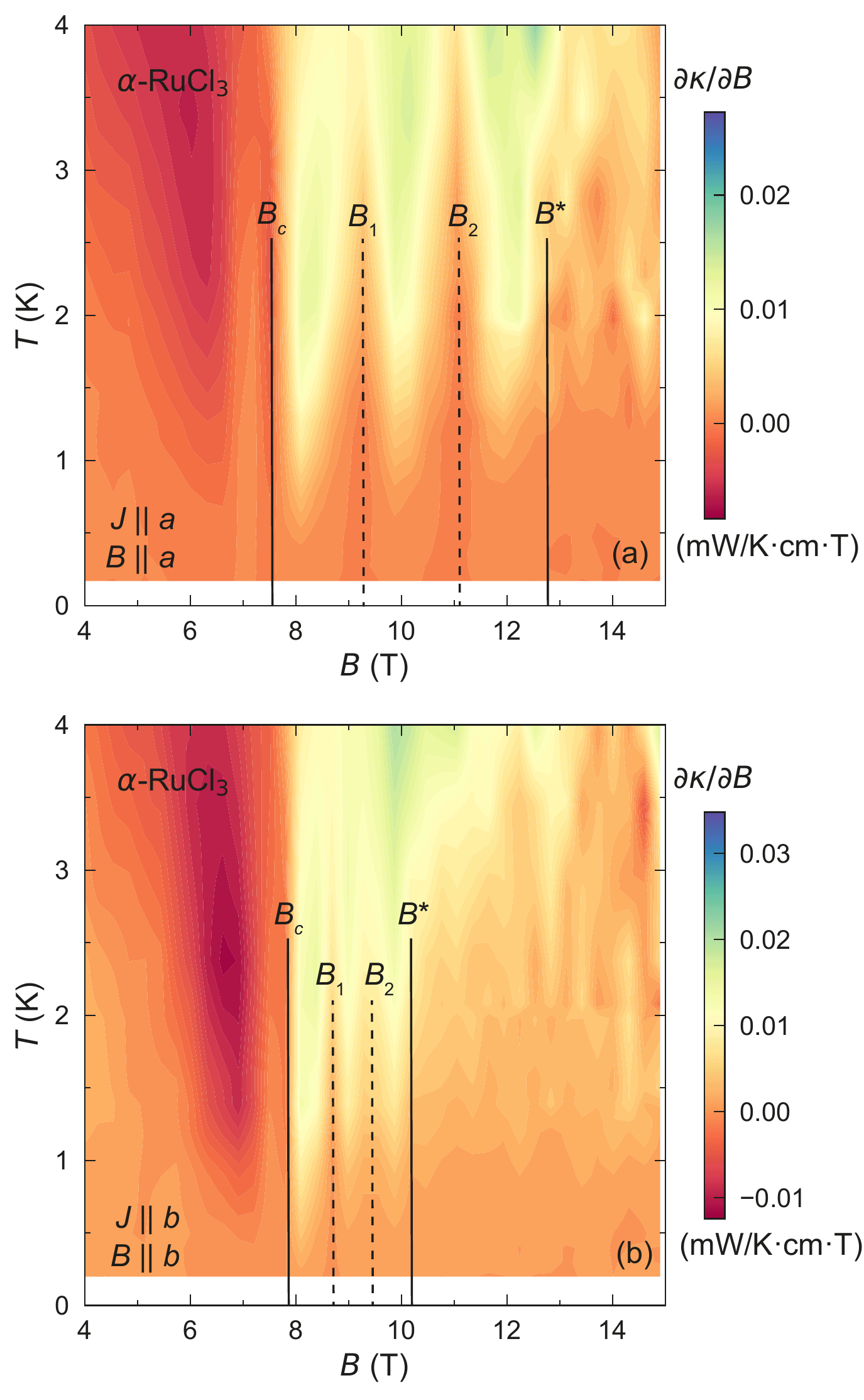}
    \caption{
    Contour plots of the derivative $\partial \kappa / \partial B$, mapped as a function of temperature and in-plane field,
    for:
    (a) $B \parallel a$, $J \parallel a$;
    (b) $B \parallel b$, $J \parallel b$.
        The vertical black lines mark the location of the four characteristic fields \Bc, $B_1$, $B_2$, and \Bstar.}
    \label{Fig6}
\end{figure}

In Fig.~\ref{Fig6}, we plot the derivative $\partial \kappa / \partial B$, obtained from our full sets of isotherms taken on our two samples (Figs.~\ref{Fig7} and \ref{Fig8}), displayed in two contour maps as a function of temperature and field,
one panel for each field direction: $B \parallel a$ and $B \parallel b$.
These contour maps clearly show how the four characteristic fields \Bc, $B_1$, $B_2$, and \Bstar~remain equally spaced in going from $B \parallel a$ to $B \parallel b$ even though the spacing shrinks by a factor of 2 or so.
This again highlights the intimate link between the oscillations ($B_1$ and $B_2$) and the transitions (\Bc~and \Bstar).


\section{DISCUSSION}

\subsection{Nature of the heat carriers}

Before we discuss the origin of the oscillations in $\kappa$ vs $B$ observed in \RuCl, let us consider the nature of the heat carriers responsible for $\kappa$.
The dependence of $\kappa$ in \RuCl~on both temperature, over the full range up to 100~K, and in-plane magnetic field, up to 18~T, is well described by assuming that the only carriers of heat are phonons and that these are scattered by spin fluctuations~\cite{hentrich_unusual_2018}. 
Such a model, involving a magnetic excitation spectrum with a gap that grows with field,
captures remarkably well the dramatic changes observed in $\kappa$ as the in-plane field is increased,
for temperatures above 7~K, the regime where no order or oscillations are observed (Fig.~\ref{Fig1}).
Phonons clearly dominate the thermal conduction in \RuCl.

\begin{figure}[t]\centering
    \includegraphics[width = 0.5\textwidth]{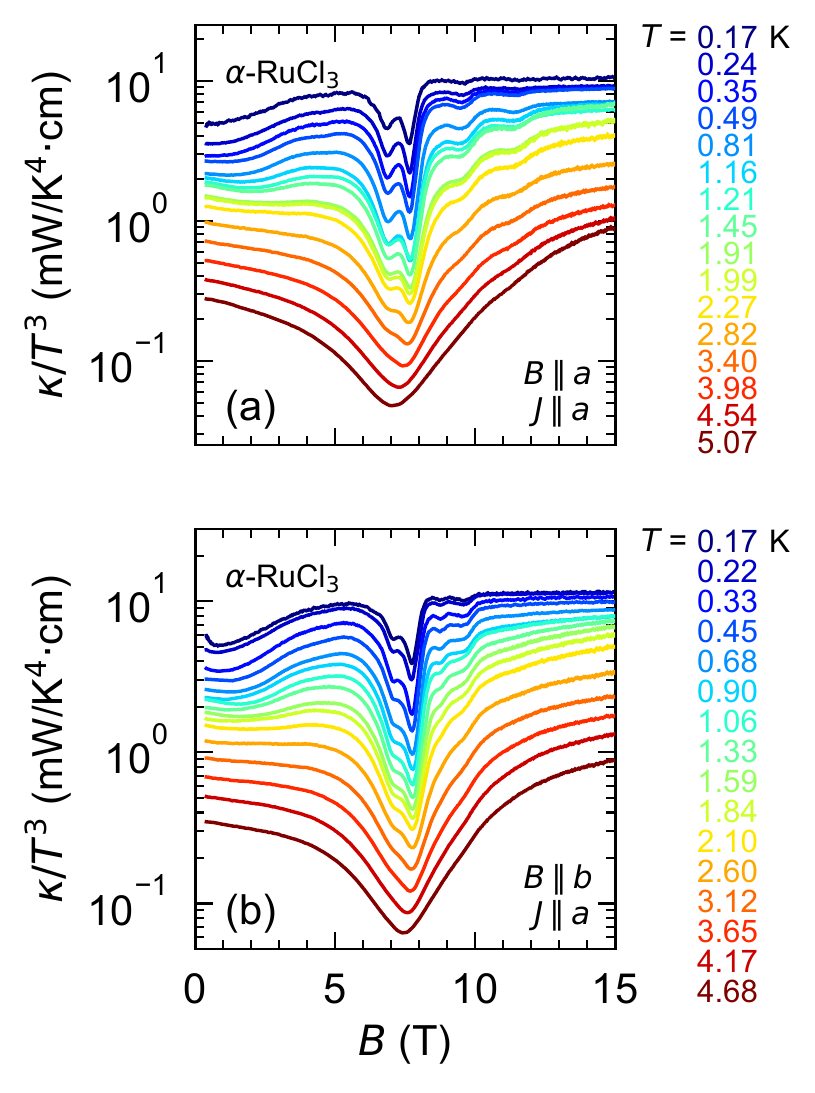}
    \caption{
    Full set of isotherms, measured on sample S1, plotted as $\kappa / T^3$ vs $B$ in a semilog plot,
    for $J \parallel a$:
    (a) $B \parallel a$;
    (b) $B \parallel b$.
        }
    \label{Fig7}
\end{figure}

\begin{figure}[t]\centering
    \includegraphics[width = 0.5\textwidth]{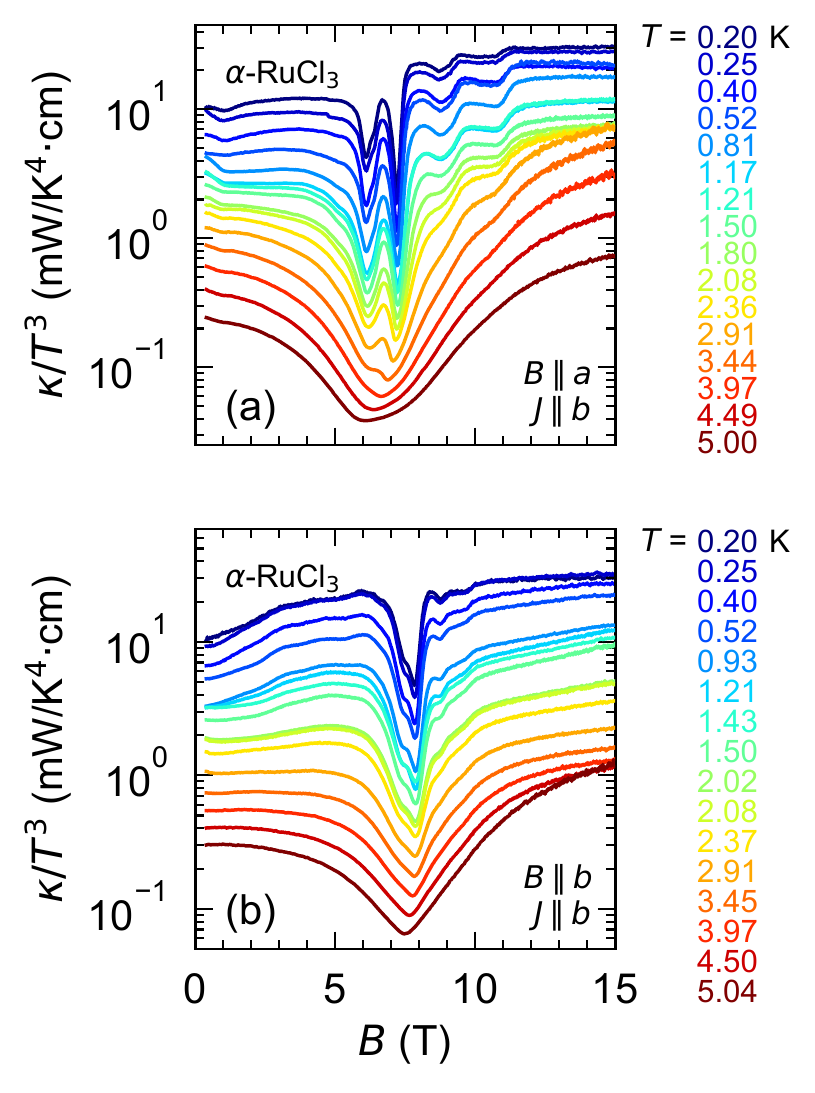}
    \caption{
    Full set of isotherms, measured on sample S2, plotted as $\kappa / T^3$ vs $B$ in a semilog plot,
    for $J \parallel b$:
    (a) $B \parallel a$;
    (b) $B \parallel b$.
        }
    \label{Fig8}
\end{figure}

What about the regime below 4~K and between \Bc~and \Bstar~where oscillations are observed?
Is there evidence here for heat carriers beyond phonons?
A recent study reports a slight anisotropy in the specific heat $C$ of \RuCl~whereby 
$C(B \parallel b) > C(B \parallel a)$ below $T \simeq 2$~K~\cite{tanaka_thermodynamic_2022}.
This anisotropy is attributed to fermions that are gapless for $B \parallel b$
and gapped for $B \parallel a$.
Suetsugu \etal~propose that a similar anisotropy is reflected in the thermal conductivity,
as they observe 
$\kappa(B \parallel b) > \kappa(B \parallel a)$ below $T \simeq 0.5$~K~\cite{suetsugu_evidence_2022},
but this is not convincing as the reverse is true at high temperature, 
\ie~$\kappa(B \parallel a) > \kappa(B \parallel b)$
at $T = 1.0$~K~\cite{suetsugu_evidence_2022}.
%
%
We also observe the same reversal of the anisotropy with increasing temperature (Figs.~\ref{Fig7},\ref{Fig8}).
These small anisotropic effects of the magnetic field on $\kappa$~are most likely coming from an
anisotropy in the magnetic scattering of phonons.

Suetsugu \etal~also emphasize the sharp jump at \Bstar~for $B \parallel a$,
and attribute it to a (first-order) topological transition, because they observe it only for $B \parallel a$ and not for $B \parallel b$.
In our own data, however, the sharp jump at \Bstar~is present for both field directions,
as seen in Fig.~\ref{Fig5}.

In summary,
there is nothing in the thermal conductivity of \RuCl~that indicates clearly
the presence of fermionic excitations,
or indeed of any excitations other than phonons.
(In the partially spin polarized state beyond the AF phase, magnons should contribute to heat transport,
but not at low temperature since they are gapped.)
%


\subsection{Origin of the oscillations}

Some theoretical studies find that a QSL state is plausible just above \Bc,
and its excitations could be gapless spinons~\cite{villadiego_pseudoscalar_2021}.
Such spinons have a Fermi surface and this could give rise to quantum oscillations.
%
%
However, two facts about the oscillations in $\kappa$ vs $B$ observed in \RuCl~(Fig.~\ref{Fig1})
immediately argue against quantum oscillations.
The first is the field direction being parallel to the layers.
Indeed, in a 2D system the Fermi surface is expected to be a cylinder with its axis normal to the layers---as
in the quasi-2D metal Sr$_2$RuO$_4$, for example~\cite{bergemann_detailed_2000}---which
would not produce quantum oscillations for $B \parallel a$ (or $B \parallel b$).
Assigning the oscillations in \RuCl~to be quantum oscillations implies that the spinon Fermi surface is
a 3D object like a sphere---difficult to imagine for a 2D system.
The second fact is the sheer magnitude of the oscillations.
In Fig.~\ref{Fig2}, we see that the oscillatory part of the conductivity can
account for as much as 50\% (peak-to-peak) of the background $\kappa$ at $T \simeq 1$~K and $B \simeq 10$~T~\cite{czajka_oscillations_2021}.
This is enormous if attributed to quantum oscillations due to fermions.
Indeed, in a metal like the cuprate YBa$_2$Cu$_3$O$_y$, for example, quantum oscillations in $\kappa$ vs $B$ coming from electrons
have a peak-to-peak amplitude of 0.2\% at $T = 1.8$~K and $B = 45$~T~\cite{grissonnanche_direct_2014}.
Note also the comparison in Fig.~\ref{Fig2} of our data on sample S1 (black curve) and the data of Czajka \etal~\cite{czajka_oscillations_2021}
(pink curve). 
Our sample has a higher conductivity, which reflects a lower level of disorder.
Yet its oscillations are smaller, whereas quantum oscillations grow exponentially with decreasing disorder (increasing mean free path of fermions).

Our anisotropy study provides a third argument against quantum oscillations from fermions.
If the minima at $B_1$ and $B_2$ are those of quantum oscillations intrinsic to a QSL state, as proposed by Czajka \etal~\cite{czajka_oscillations_2021}, why are there only two such minima (above \Bc)?
The authors postulate that the QSL phase ends at \Bstar, in accordance with the absence of further oscillations at higher field.
Now the period of quantum oscillations is an intrinsic property of the QSL,
imposed by the volume of the Fermi surface of those neutral fermions (\eg{} gapless spinons).
The fact that two periods happen to fit in the interval created by the two phase transitions
at \Bc~and \Bstar, say for $B \parallel a$, is necessarily an accident.
If we then change the field direction to $B \parallel b$, the period of quantum oscillations
will change according to the topology of the spinon Fermi surface, which has nothing to do with 
the critical fields \Bc~and \Bstar.

Yet what we observe, as illustrated in Fig.~\ref{Fig5}, is that the period of the oscillations in \RuCl~changes in proportion
to the change in the separation between the two critical fields \Bc~and \Bstar.
There are only two oscillations for both field directions, even though the interval between \Bc~and \Bstar~is twice as large for $B \parallel a$.
We conclude that the oscillations are intimately related to \Bstar.
More specifically, \Bstar~marks the end of a transition centered at $B_2$.
In other words, the minima at $B_1$ and $B_2$ mark two additional magnetic transitions above those at $B_0$ and \Bc, as proposed by Bruin \etal~\cite{bruin_origin_2022} on the basis that additional anomalies are also detected in the magnetization, at the same field values,
and these can be tracked to transitions vs temperature in zero field, lying above \TN.

The nature of the transitions at $B_1$ and $B_2$ remains to be elucidated.
The fact that the values of $B_1$ and $B_2$ are the same in different samples~(Fig.~\ref{Fig2}),
from different growth methods, suggests that they are generic features of \RuCl.
However, the weakness of the associated specific heat anomalies suggests that
the magnetic structures involved may not occupy the full volume of the sample and 
could be nucleated by local defects in the crystal structure, such as stacking faults.

%


\section{SUMMARY}

%
We investigated the proposal that oscillations detected in the thermal conductivity $\kappa$ of the Kitaev material \RuCl~as a function of magnetic field $B$ are quantum oscillations from neutral fermions that are the emergent excitations of a quantum spin liquid (QSL) state, which would exist in a regime of in-plane magnetic fields just above its antiferromagnetic (AF) phase, and below some critical field \Bstar, 
as proposed by Czajka and coworkers~\cite{czajka_oscillations_2021}.
We have measured the thermal conductivity $\kappa$ of \RuCl~as a function of field up to 15~T at several temperatures between $T = 0.2$~K and $T = 5$~K for the two in-plane field directions $B \parallel a$ and $B \parallel b$.
For both field directions, we observe two oscillations in $\kappa$ vs $B$ contained between \Bc, the critical field where the AF phase ends, and \Bstar, the threshold field above which no further oscillations are seen (the putative critical field where the QSL phase ends). 
We find that as the field is changed from $B \parallel a$ to $B \parallel b$, the interval between the two transition fields \Bc~and \Bstar~shrinks by a factor of 2, and so does the period of the two oscillations contained in that interval. 
Because there is no reason for quantum oscillations---dictated by the Fermi surface of fermions---to be related to the critical field of the QSL phase (\Bstar), we argue that the correlation between oscillation period and field interval (\Bstar~$-$~\Bc) is instead evidence that the oscillations are produced by secondary magnetic transitions similar to the main transition at \Bc. \\
%

%
Our conclusion is consistent with that of Bruin \etal~\cite{bruin_origin_2022} and with
theoretical works that report the absence of an intermediate field-induced region,
in the phase diagram as a function of in-plane field (Fig.~\ref{Fig1}), 
that could potentially harbor a QSL state~\cite{gordon_theory_2019,winter_probing_2018}.


\section*{ACKNOWLEDGMENTS}

We thank S. Fortier for his assistance with the experiments. 
L. T. acknowledges support from the Canadian Institute for Advanced Research (CIFAR) as a CIFAR Fellow 
and funding from the Institut Quantique, the Natural Sciences and Engineering Research Council of Canada (NSERC; PIN:123817), 
the Fonds de Recherche du Qu\'ebec - Nature et Technologies (FRQNT), the Canada Foundation for Innovation (CFI), 
and a Canada Research Chair. 
This research was undertaken thanks in part to funding from the Canada First Research Excellence Fund.
Work at the University of Toronto was supported by NSERC (RGPIN-2019-06449 and RTI-2019-00809), CFI, 
and the Ontario Ministry of Research and Innovation.






%

\end{document}